# Initiation of the Worthington jet on the droplet impact

Ken Yamamoto[a,*], Masahiro Motosuke[a], Satoshi Ogata[b]

[a] Tokyo University of Science, 6-3-1 Niijuku, Katsushika-ku, Tokyo 125-8585, Japan
[b] Tokyo Metropolitan University, 1-1 Minami-Osawa, Hachioji, Tokyo 192-0397, Japan

[*] Corresponding author.
E-mail address: yam@rs.tus.ac.jp
Tel.: +81 (0)3 5876 1718

**Abstract**

Deformation of liquid droplets by impingement induces Worthington jet in certain range of the impact velocity. Although the growth of the jet as well as its tip velocity is predicted from similar cases to the droplet impact, the mechanism of the jet generation is yet to be understood. In this study, high-speed visualization of the droplet impact on a superhydrophobic surface was conducted to understand the initiation of the jet generated by a collapse of an air cavity. Water droplets whose diameter are 2.0 mm and 3.0 mm were generated and the Weber number of the droplet was varied in a range of 2 to 20. The jet velocity was measured from the captured images and it was found that it has a peak at the Weber number equals to approximately seven. The resulting jet velocity at the peak was approximately 15 folds higher than the impact velocity. Moreover, we observed that surface waves were generated by a collision of the droplet on the solid and the waves induced an oscillation of the droplet cap as they propagate from the solid–liquid contact line to the top portion of the droplet. Furthermore, we found the phase of the oscillation is related to the Weber number and it influences the jet velocity significantly, as it determines the initial condition of the jet generation.

Wetting transition in the droplet impact is an issue of interest in many industrial fields because the droplet behavior changes significantly at the transition point. In the simplest scenario, this transition can be predicted from a balance between the inertia and surface tension.[1–6] However, it was found recently that the transition occurs at lower Weber number, $We$ (= $\rho U_{jet} D_0/\gamma$, $\rho$, $U_0$, $D_0$, and $\gamma$ are liquid density, impact velocity, initial droplet diameter, and surface tension, respectively) than expected from the balance, if the droplet generates a liquid jet during the impact.[7,8] It was also found that the jet was generated in a limited range of $We$ (typically of the order of $10^0$–$10^1$). However, the detailed mechanism of the jet generation as well as the $We$-dependency is yet to be investigated.

Generation of a jet from a free surface is observed in many situations such as pinch off of a droplet from a nozzle,[9] impact of an object onto a liquid surface,[10–14] breakup of a bubble at the liquid–gas interface,[15,16] and oscillation of interfaces.[17,18] The jet generated by the droplet impact is sorted as the Worthington jet, of which mechanism is explained by a collapse of an air cavity formed at the center of the deformed droplet[19,20]. The typical diameter of the jet is tens or hundreds of micrometer and the jet velocity $U_{jet}$ reaches close to or more than 10 folds of the impact velocity of the droplet.[19,21–23] The generation and growth of the Worthington jet is modeled as a shrinking conical air cavity[12,13,17,24]: the cavity is axisymmetrically squeezed and collapsed. At the collapse point, the radial velocity of the liquid abruptly diminishes and an axial velocity component is generated to conserve the volume. The resulting jet velocity in the droplet impact can be roughly estimated as a function of the impact velocity with a given jet radius[8,19]. However, the model does not precisely explain a generation and size of the initial rise of the jet tip, which influence the jet velocity significantly.[13] Furthermore, empirical observation of the initial rise is quite challenging due to the optical distortion caused by deformed interfaces of the droplet as well as the small spatial and temporal scale.[19]

In this study, the jet generation process was visualized from two different view angles with an aid of a high-speed camera to better understand the jet generation mechanism. We measured $U_{jet}$ from the captured images and relationships between $U_{jet}$, jet generation, and impact velocity were discussed.

The experimental setup for this study was almost the same as the one employed in the previous study[8] and therefore only brief description will be given here. The test surface was prepared by aligning stainless razor blades (Hi-Stainless Platinum Double Edge Razor Blades, FEATHER Safety Razor, averaged blade angle of ~13°) in parallel. The width between each blade tip $w$ was set to approximately 80 μm. The apparent contact angle of the test surface was measured as approximately 150°. We chose distilled water as the working liquid. The droplets were produced by infusing the liquid from a needle connected to a syringe pump (NEXUS 6000, Chemyx Inc.) with a flow rate of 0.01 mL/min. The system can stably produce droplets with constant diameter $D_0$, which can be varied by selecting the needle: a 34G flat-tipped needle for $D_0$ = 2.0 mm and an 18G needle for $D_0$ = 3.0 mm, respectively. The droplet impact velocity $U_0$ was controlled by releasing the droplets from different height and the sequences of the collision were recorded from horizontal and obliquely above views at 4000–15000 fps by a high-speed camera (VW-9000,

KEYENCE) combined with a long-distance zoom lens (VH-Z35, KEYENCE) under the backlight illumination by a LED light source (LA-HDF6010WD, Hayashi Watch-works). The experimental range of *We* was from 2 to 20.

A typical Worthington jet generated from a deformed droplet is shown in Fig. 1. The generation of the thin jets was observed only in a certain range of *We* by the following reasons: the lower limit (*We* ~ 5) is explained by less deformation of the droplet due to less inertia, whereas the upper threshold is given by the wetting transition due to the inertia:[8] the wetting transition occurs when the dynamic pressure $P_d$ (= $0.5\,\rho U_0^2$) exceeds the Laplace pressure of the pinned meniscus located at edges of the solid surface. Because the local Laplace pressure $P_L$ can be estimated as $\sim \gamma/r_m$, the critical velocity $U_c$ can be obtained from a balance of these pressure as $U_c \sim [\gamma/(\rho r_m)]^{1/2}$, where $r_m$ is the radius of the pinned meniscus that can be obtained from the surface geometry and the advancing contact angle $\theta_a$ as $r_m \sim -w/\cos\theta_a$. In our case ($w$ = 80 μm, $\theta_a$ = 124°), we obtain $U_c \sim 0.67$ m/s, or the critical Weber number $We_c$ of approximately 12 and the actual wetting transition occurred at $We \sim 10$. Note that the liquid jets were also observed for $We > 10$, but we distinguished them because the symmetry of the interface was broken due to the wetting and it made their typical radius much larger than those observed in $We < 10$.

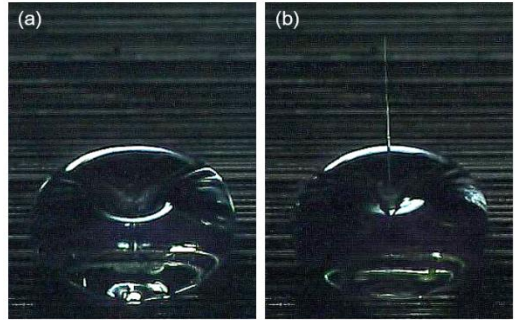

**FIG. 1** Worthington jet generated during the droplet impact ($D_0$ = 2 mm, *We* = 6.2). (a) 4 ms after the impact. The rim of the droplet is recoiling and the air cavity is formed at the center of the deformed droplet. (b) 4.25 ms after the impact. The air cavity is squeezed and the Worthington jet is generated.

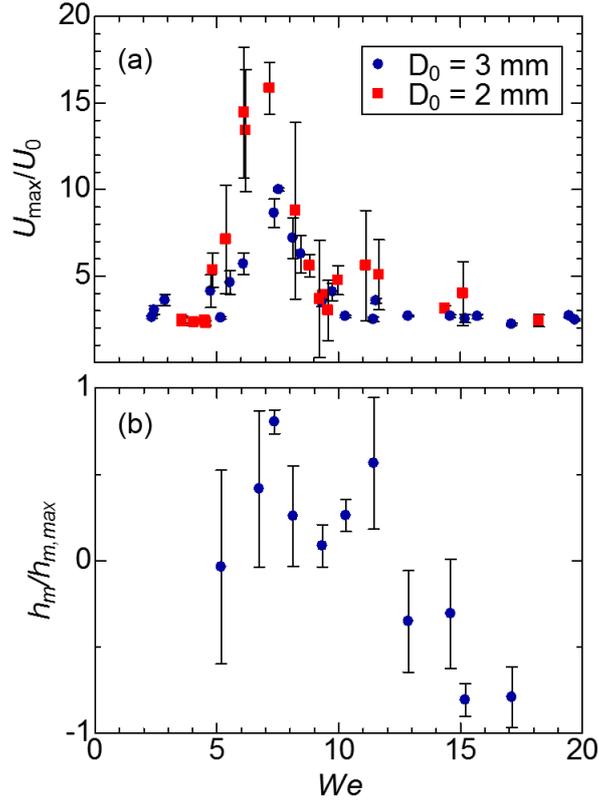

**FIG. 2** (a) Dependency of $U_{max}/U_0$ as a function of $We$. $U_{max}$ indicates either $U_{jet}$ or the maximum velocity of the droplet surface at the highest point. The jets were observed at $We > 5$. The wetting transition by inertia occurred at $We \sim 10$. (b) Phase of the droplet cap shape immediately before the collapse of the cavity as a function of $We$ ($D_0 = 3.0$ mm). The phase $h_m/h_{m,\max} = 1$, 0, and −1 correspond to convex, flat, and concave shapes of the cap, respectively.

To evaluate characteristics of the jet generation, the jet velocity from horizontal view was measured. Figure 2a shows measured velocity as a function of $We$. Note that we measured either the jet velocity ($We > 5$) or the maximum velocity of the droplet cap ($We < 5$), and the velocity is displayed as $U_{max}$ in a normalized form. Below the lower limit of the jet generation ($We < 5$), $U_{max}$ is few times higher than the impact velocity due to the surface deformation. In the jet generation range, $U_{max}/U_0$ increases with $We$-, indicates the peak at $We \sim 7$ and decreases at $7 < We < 10$. This trend corresponds to the literature data[19,22,23] except for the region of $We > 10$. The discrepancy in $We > 10$ is supposedly due to the wetting transition: the droplet shape was varied when the wetting transition occurred, and it prevented to squeeze the air cavity properly (this effect might be enhanced substantially by the anisotropic surfaces employed in this study.) For $We$ larger than 10, the droplets still generate the jets despite the wetting transition. However, the jet radius is much larger than that in $We = 5$–10, which results in lower $U_{max}/U_0$.

The results shown in Fig. 2a suggests the jet velocity cannot simply be determined from the squeezing velocity ($\sim U_0$) with an assumption of a constant radius of the initial jet tip. Because it implies the cavity shape varies with the impact velocity, we investigated effects of surface waves generated by the impact[25], whose wavelength is expressed as $\sim[\gamma/(\rho U_0^2)]^{0.5}$.[9,18] Figure 3 shows a time series of an impacting droplet. One can observe the surface waves are generated from the contact line and propagated towards the cap of the droplet and the droplet forms a pyramidal shape.[25] The primary wave reaches the cap at ~2 ms after the impact and the cap fluctuates periodically. The amplitude of the fluctuation increases by a superposition of the waves while the radius of the cap gradually decreases. As a result, the fluctuation frequency of the cap decreases with the elapsed time and it is of the order of $10^2$ Hz at the last moment before the cavity collapse. This implies the local shape of the cavity near the axis is largely different in a timescale of millisecond, which is close to the timescale of the squeezing. Here, we assume that the cap height from the local free surface in the $n$th fluctuation $h_n$ can be expressed as $h_n = A_n\cos(\omega_n t + \varphi_n)$, the maximum height $h_{n,\max} = A_n$, and the count of the fluctuations before collapse is $n = m$. By employing $t_1$ and $t_2$, which represent times at $h_m/h_{m,\max} = 1$ and $-1$, the phase of the $m$th fluctuation at the moment of the collapse can be estimated as $(t_c - t_1)/[2(t_2 - t_1)]$, where $t_c$ is the collapse time.

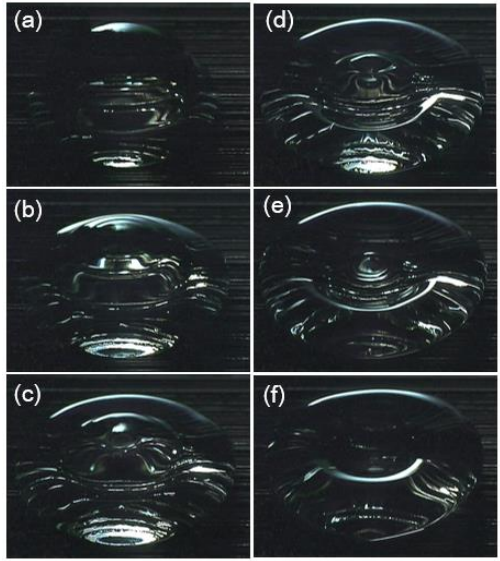

**FIG. 3** Successive images of the impacting droplet ($D_0 = 3.0$ mm) at (a) 1.50 ms, (b) 2.25 ms, (c) 3.00 ms, (d) 3.75 ms, (e) 4.50 ms, and (f) 6.00 ms after the impact. Surface waves are generated from the contact line and propagated toward the droplet cap.

The measured phase as a function of $We$ is shown in Fig. 2b. The diagram shows the phase at the impact has a dependency of $We$, with two peaks (at $We \sim 7$ and $\sim 12$). Note that we measured the phase only for the case of $D_0 = 3$ mm owing to the limitation of the time resolution in the measurement. Recalling that the time from the impact to the cavity collapse is constant with given $D_0$,[8] the phase shift by $We$ can be considered as a result of different wavelength and its propagation velocity, which are related to the impact velocity. Although we could not observe the surface shape at the axis immediately before the collapse due to the existence of the recoiling rim, these peaks imply the shape at the axis varies in between flat (*We* ~ 5 and ~9) and convex (*We* ~ 7 and ~12) as schematically shown in Fig. 4. Moreover, the correspondence of the Weber number for the first peak and that for the peak of $U_{max}/U_0$ (Fig. 2a) implies the jet velocity depends on the shape of the free surface at the cavity axis and it reaches maximum when the shape is convex, at which, presumably, the initial jet tip radius takes minimum. In addition, although the second peak (*We* ~ 12) of the jet velocity was not observed in this study, it will be observed if we employ smaller *w*, which increases the critical Weber number, because the second peak was observed in previous studies at the corresponding velocity.[19,23] The periodicity of the peak might be derived from the increase of the number of the fluctuation *m* with the impact velocity.

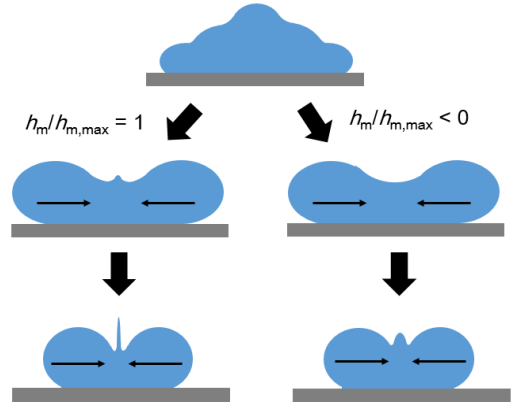

**FIG. 4** Schematic of the droplet shapes during the impact. The droplet forms pyramidal shape in early stage of the impact due to the surface wave. The droplet shape then becomes toroidal and its rim recoils toward the center. In the last stage (immediately before the jet generation) an air cavity is formed. The surface shape at the axis and the resulting jet velocity depends on the phase $h_m/h_{m,max}$. (Left) The minimum jet radius is obtained at $h_m/h_{m,max} = 1$, at which the oscillating droplet cap becomes a tiny bump that leads the smallest jet radius. (Right) The tiny bump is not generated for $h_m/h_{m,max} < 0$. In this case, the jet radius is larger than the case of $h_m/h_{m,max} = 1$.

In summary, the generation process of the Worthington jet from water droplets (2.0 mm and 3.0 mm in diameter) impacted on the superhydrophobic surface was visualized from horizontal and obliquely above views to understand the jet generation mechanism. The experiment was performed with various impact velocities ranged from 0.23 m/s to 0.81 m/s (equivalent to 2 < *We* < 20). The jet velocity was measured from horizontal images and we found it has a peak at *We* ~ 7, in contrast to a simple prediction with a universal cavity shape. The other visualization angle revealed that this discrepancy was derived from the effect of the surface waves generated by the impact: it generates oscillations of the droplet cap and the magnitude of the jet velocity is determined by the matching of the phase of the oscillation and squeezing of the air cavity at the center.

**Acknowledgment**

This study was financially supported by Japan Society for the Promotion of Science (KAKENHI Grant No. 22760134).